\documentclass[conference, 9pt]{IEEEtran}
\IEEEoverridecommandlockouts

\usepackage{cite}
\usepackage{amsmath,amssymb,amsfonts}
\usepackage{algorithmic}
\usepackage{graphicx}
\usepackage{textcomp}

\usepackage{cancel}
\usepackage{multirow}
\usepackage{amsmath,graphicx}
\usepackage{cite}
\usepackage{booktabs}
\usepackage{hyperref}

\usepackage{color,multirow,amsfonts,amssymb,float,enumitem}
\usepackage{bibspacing}

\usepackage{bbding}
\usepackage{pifont}
\newcommand{\cmark}{\ding{51}}
\newcommand{\xmark}{\ding{55}}

\newcommand{\whitediamond}{\mathord{\diamond}}

\newcommand{\hnxu}[1]{{\color{black}{#1}}}

\usepackage[table]{xcolor}

\def\BibTeX{{\rm B\kern-.05em{\sc i\kern-.025em b}\kern-.08em
    T\kern-.1667em\lower.7ex\hbox{E}\kern-.125emX}}
    
\begin{document}
\bstctlcite{IEEEexample:BSTcontrol}

\title{Effective and Efficient Mixed Precision Quantization of Speech Foundation Models
}

\author{
\IEEEauthorblockN{Haoning Xu$^1$, Zhaoqing Li$^1$, Zengrui Jin$^1$, Huimeng Wang$^1$, Youjun Chen$^1$, Guinan Li$^1$,\\Mengzhe Geng$^2$, Shujie Hu$^1$, Jiajun Deng$^1$, Xunying Liu$^1$}
\IEEEauthorblockA{
\textit{$^{1}$The Chinese University of Hong Kong, Hong Kong SAR, China};  
\textit{$^2$National Research Council Canada, Canada} }
\texttt{\{hnxu, xyliu\}@se.cuhk.edu.hk}

}

\maketitle


\begin{abstract}
This paper presents a novel mixed-precision quantization approach for speech foundation models that tightly integrates mixed-precision learning and quantized model parameter estimation into one single model compression stage. Experiments conducted on LibriSpeech dataset with fine-tuned wav2vec2.0-base and HuBERT-large models suggest the resulting mixed-precision quantized models increased the lossless compression ratio by factors up to 1.7x and 1.9x over the respective uniform-precision and two-stage mixed-precision quantized baselines that perform precision learning and model parameters quantization in separate and disjointed stages, while incurring no statistically word error rate (WER) increase over the 32-bit full-precision models. The system compression time of wav2vec2.0-base and HuBERT-large models is reduced by up to 1.9 and 1.5 times over the two-stage mixed-precision baselines, while both produce lower WERs. The best-performing 3.5-bit mixed-precision quantized HuBERT-large model produces a lossless compression ratio of 8.6x over the 32-bit full-precision system.

\end{abstract}


\section{Introduction}

In recent years, self-supervised learning (SSL) based speech foundation models such as wav2vec2.0~\cite{baevski2020wav2vec}, HuBERT~\cite{hsu2021HuBERT} and WavLM~\cite{chen2022wavlm} have demonstrated performance advancements across a range of applications such as automatic speech recognition (ASR). 
However, the practical deployment of current speech foundation models to on-device and resource-constrained scenarios is hindered by their memory footprint and computational cost.

To address this issue, neural network model compression techniques have been widely studied including, but not limited to: \textbf{1) architecture compression} methods that aim to minimize model structural redundancy using weight pruning~\cite{pru3,pru5,jiang2023accurate}, low-rank matrix factorization~\cite{lr3,li2023lossless} and knowledge distillation~\cite{rathod2022multi,park2023conformer}; and \textbf{2) low-bit quantization} approaches that reduce memory footprint by replacing floating point weights with low precision values~\cite{uq-2bcfm,uq-4bcfm,ibert,zhao2024ada}. 

Model compression research in the context of SSL speech foundation models to date focuses on architectural compression using either weight pruning~\cite{lai2021parp,hj,lodagala2023pada}, knowledge distillation~\cite{distilHuBERT,dist1,dist2,dist3,chang2023colld,cho2023sd,de2023distilling,wang2022lightHuBERT}, or both~\cite{peng2023dpHuBERT}. Only a few recent studies have been conducted on speech foundation model quantization \cite{mp-w2v,onepass-zq,usmlite,mq-person,peng2021shrinking} which primarily exploit uniform-precision quantization (i.e., all quantized parameters are compressed to identical bit-widths)~\cite{mp-w2v, onepass-zq, usmlite,peng2021shrinking}. 
Furthermore, larger model compression ratios can be achieved by combining both architectural compression and low-bit quantization~\cite{li2023lossless,onepass-zq,mp-w2v,usmlite}. 

However, these prior studies suffer from the following limitations: \textbf{1) Uniform-precision quantization} fails to account for the fine-grained and locally varying performance sensitivity to quantization at different model internal components~\cite{uq-2bcfm,mp-w2v,usmlite,peng2021shrinking}. In this direction, some methods require manual, hand-crafted assignment of layer-level quantization bit-widths~\cite{tan2023dqmix,zhen2023sub8cfm}. Hence, more powerful mixed-precision quantization approaches~\cite{mq-jh, mq-person} that automatically determine the optimal, locally varying quantization precision settings are preferred. \textbf{2) Inconsistency between precision learning and quantized parameter estimation} creates two separate and disjointed stages during system compression and leads to large performance degradation~\cite{mq-person}.
\textbf{3) Inefficiency of two-stage quantization} approaches ~\cite{mq-jh, zhao2021automatic} further increases the overall system compression time in addition to post-quantization performance loss.
\textbf{4) Significant performance degradation} is often observed in terms of ASR WER increase after performing quantization~\cite{mp-w2v,kim2022integer,yao2022zeroquant,cai2020zeroq}. It is important to note that most studies failed to clearly define the criterion, such as statistical significance tests\cite{gillick1989some}, to differentiate "acceptable" and "unacceptable" performance loss due to quantization. 

To this end, this paper presents a novel mixed-precision quantization approach for SSL speech foundation models that tightly integrates mixed-precision learning and quantized model parameter estimation into one single model compression stage. Neural architecture search (NAS)~\cite{liu2018darts} based automatic mixed-precision learning is performed over multiple weight-sharing systems of different quantization bit-widths. KL regularization is further applied to mitigate the performance degradation caused by quantization\cite{onepass-zq}, especially for systems with ultra-low-precision bid-width (e.g., 2-bit). 
Experiments conducted on Librispeech dataset with fine-tuned wav2vec2.0-\textit{base} and HuBERT-\textit{large} models suggest the resulting mixed-precision quantized models increased the lossless compression ratio by factors up to 1.7x and 1.9x over the respective uniform-precision and two-stage mixed-precision quantized baselines that perform precision learning and model parameters quantization in separate and disjointed stages, while incurring no statistically word error rate (WER) increase over the 32-bit full-precision models. The system compression time of wav2vec2.0-\textit{base} and HuBERT-\textit{large} models is reduced by up to 1.9 and 1.5 times over the two-stage mixed-precision baselines, while both produce lower WERs. The best-performing 3.5-bit mixed-precision quantized HuBERT-\textit{large} model produces a lossless compression ratio of 8.6x over the 32-bit full-precision system.


Our proposed approaches achieve the following improvements: 

\textbf{1)} Compared with the uniform-precision quantized wav2vec2.0-\textit{base} and HuBERT-\textit{large} systems, our compressed mixed-precision quantized models boosted the \textit{lossless}\footnote{Lossless in this paper refers to no statistically significant WER increase against the 32-bit full-precision models on the test set.} compression ratios from 3.7x to 4.7x and from 4.5x to 6.8x, respectively.

\textbf{2)} In contrast to the two-stage mixed-precision systems:
the compression time of the 4-bit and 3.8-bit mixed-precision quantized wav2vec2.0-\textit{base} and HuBERT-\textit{large} systems are heavily reduced by maximum factors of 1.9x and 1.5x, respectively, while both demonstrate lower WERs with absolute WER reductions up to 0.55\%;


The main contributions of this paper are three-folded:

    \textbf{1)} To the best of our knowledge, this is the first work to jointly perform mixed-precision learning and quantized model training for SSL ASR systems.
    Previous mixed-precision quantization studies on SSL speech foundation models quantize the system only after an earlier, separate stage of quantization sensitivity measurement and precision setting~\cite{mq-person}, which leads to the inconsistency between precision learning and quantized parameter estimation.

    \textbf{2)} Our proposed approach reduced the overall system compression time over related prior works: 
    \textbf{a)} In contrast to prior approaches conducted on text-based BERT~\cite{zhao2021automatic}, our post-quantization fine-tuning strategy eliminates the need of re-estimating quantized parameters from scratch; \textbf{b)} Differing from the two-stage mixed-precision quantization approach~\cite{mq-jh} for Transformer LMs, our method bypasses the need of training multiple uniform-precision quantized models prior to learning the mixed-precision bit-widths.

    \textbf{3)} The obtained quantized 4.6-bit wav2vec2.0-\textit{base} system (with 8-bit CNNs\footnote{For simplicity, we use "CNNs" to denote the components of the model structure excluding the Transformer encoder.}) and 3.5-bit HuBERT-\textit{large} system (with 8-bit CNNs) achieved a maximum lossless compression ratio of \textbf{6.4x} and \textbf{8.6x}, respectively, with no statistically significant WER increase on LibriSpeech's~\cite{panayotov2015librispeech} test set.

\section{wav2vec2.0 and HuBERT Foundation Models} \label{structure}

Speech SSL models such as wav2vec2.0~\cite{baevski2020wav2vec}, HuBERT~\cite{hsu2021HuBERT}, and WavLM~\cite{chen2022wavlm} share similar Transformer backbones with supervised models. For example, wav2vec2.0 consists of a CNN encoder, a Transformer encoder, a projection layer and a code embedding layer. Transformer encoder accounts for over 90\% of the total number of parameters, where each encoder layer contains an MHSA module and an FFN module. In this paper, we fine-tune the pre-trained wav2vec2.0-\textit{base} and HuBERT-\textit{large} models with a pure CTC decoder.

In this paper, we perform quantization on the Transformer encoder (excluding the other parts) for wav2vec2.0 models and HuBERT models. To explore the upper limit of compression ratio, we also conduct experiments quantizing the other parts including a CNN encoder, a projection layer and a code embedding layer (uniformly represented by CNNs in this paper) with 8 bits, which is usually ignored by previous studies.

\begin{figure*}[ht]
    \centering
    \includegraphics[scale=0.107]
    {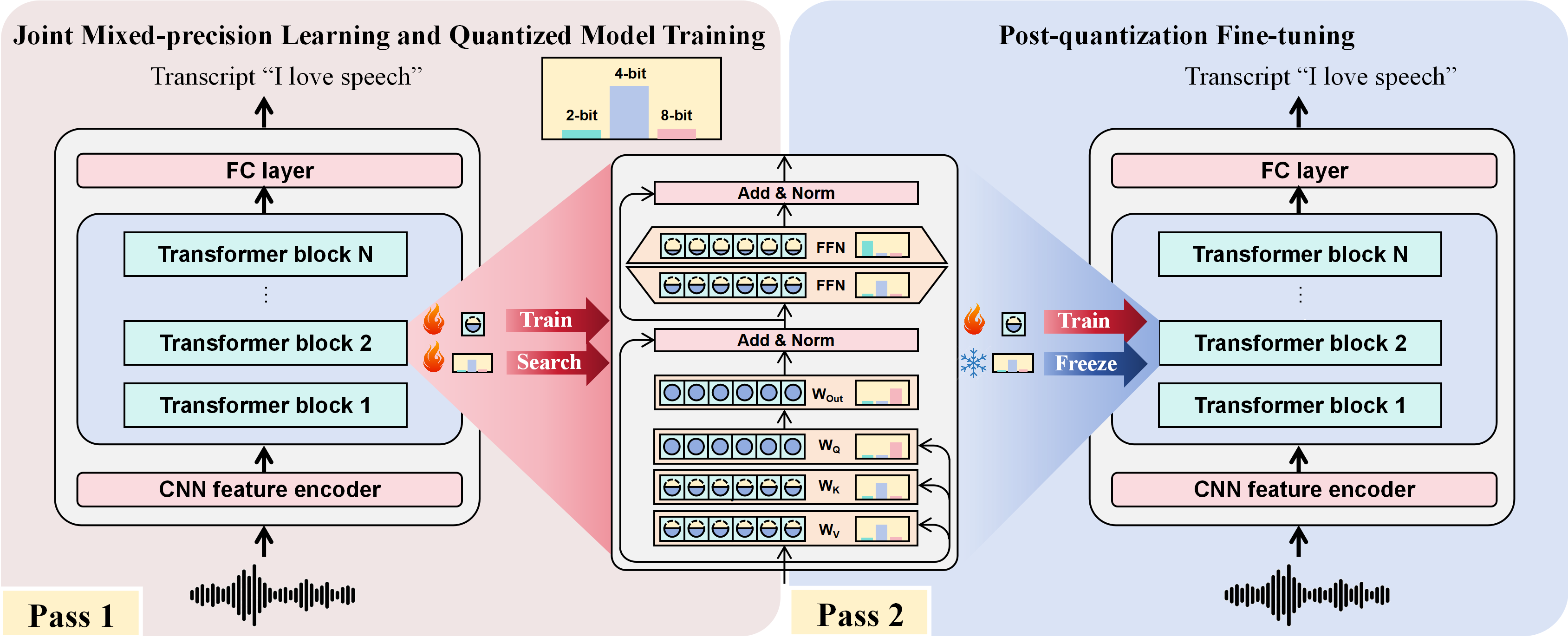}
    \vspace{-0.3cm}
    \caption{Diagram of joint mixed-precision learning and quantized model parameter estimation system with automatic mixed-precision learning over multiple weight-sharing systems of different quantization bit-widths. \textbf{Joint mixed-precision learning and quantized model training stage (Pass 1)}  : all the candidate networks of different bit-widths (i.e., 2-, 4-, and 8-bit) and the mixed-precision bit-widths are simultaneously trained and updated.  \textbf{Post-quantization fine-tuning stage (Pass 2)}: parameters can be further fine-tuned while keeping the mixed-precision bit-widths frozen.}
    \label{fig1}
    \vspace{-0.6cm}
\end{figure*}

\section{mixed-precision learning using Neural architecture search} \label{pass 1}

In this section, we present a variant form of the differentiable neural architecture search (DARTS) method, which is adapted for mixed-precision learning~\cite{liu2018darts,hu2020dsnas}.
Instead of training and storing different bit-width quantized systems separately~\cite{mq-jh}, we train a supernet that contains various weight-sharing candidates of different bit-widths simultaneously, with each being assigned a learnable parameter to measure its importance. Specifically, the $l$-th layer's output $\mathbf{h}^l$ can be computed as follows in a DARTS supernet:

\begin{equation}
\mathbf{h}^l=\sum_{i=2,4,8}\lambda_i^l\phi_i^l(\mathbf{h}^{l-1}),
\end{equation}

where $\phi_i^l(\cdot)$ denotes quantizing modules of the $l$-th layer to $i$-bit and $\lambda_i^l$ is the parameter that measures the importance of the $i$-bit quantized candidate of the $l$-th layer.

\vspace{-0.2cm}
\subsection{Gumbel-Softmax DARTS}
In order to produce approximately a one-hot vector for selecting candidates, a Gumbel-Softmax\cite{maddison2022concrete} is used to sharpen the distribution of the importance parameters.
Specifically, different from the Softmax function which directly normalizes the output, Gumbel-softmax is obtained by gumbel sampling the output of the Softmax function, which is given by:
\vspace{-0.2cm}
\begin{equation}
    \lambda_i^l=\frac{\exp((\log\alpha_i^l+G_i^l)/T)}{\sum_{j=2,4,8}\exp((\log\alpha_j^l+G_j^l)/T)},
\end{equation}
where $\alpha_i^l$ is an architecture-dependent parameter determining their contribution during NAS search. $G_i^l = -\log(-\log(U_i^l))$ is the Gumbel variable, and $U_l^i$ is a uniform random variable. When the temperature parameter $T$ approaches 0, the Gumbel-Softmax distribution is close to a categorical distribution.
\vspace{-0.2cm} 
\subsection{Joint mixed-precision learning and quantized model training}
\hnxu{Traditional two-stage mixed-precision methods~\cite{mq-jh} require separately training multiple uniform-precision (e.g., 2-, 4-, 8-bit) quantized systems first, followed by an independent measurement of sensitivities of different layers to search for the precision bit-widths, and then a mixed-precision quantized system is trained from scratch using the searched mixed-precision bit-widths. This can lead to inconsistency between mixed-precision learning and quantized model training. In this paper,} the proposed framework, taking the $l$-th layer as an example, concurrently optimizes the importance distribution $\lambda_i^l$ and the weights of each candidate $\phi_i^l(\cdot)$, which enables the model to identify the optimal mixed-precision bit-widths for quantization while maximizing its performance, thereby mitigating the mismatch between mixed-precision learning and quantized model training. This process is referred to as \textbf{joint mixed-precision learning and quantized model training (Pass 1)}. 

Finally, we design the training loss as:

\begin{equation}
    \mathcal{L} =\mathcal{L}_{mp} +\eta\cdot\mathcal{C}_{size},\label{Pass 1 eq}
\end{equation}
where $\mathcal{L}_{mp}$ is the CTC loss of the mixed-precision quantized system with the real-time searched mixed-precision bit-widths, $\mathcal{C}_{size}$ represents the complexity penalty term, expressed as the exact model size of the mixed-precision quantized system with a certain mixed-precision bit-widths, and $\eta$ is a constant coefficient to control the overall model bit-width.

\section{Kullback-Leibler Divergence Regularization}
\hnxu{Inspired by~\cite{onepass-zq}, 
KL regularization is exploited in this paper so that the full-precision speech foundation model (teacher) can better guide the lower-bit-width quantized systems (students) during compression. 
In addition to only guiding the mixed-precision quantized system, we add extra 3 sub-systems (i.e., 2-bit, 4-bit, and 8-bit uniformly quantized systems) also as students. When integrated into the same quantization cycle, the above can consistently benefit both mixed-precision learning and quantized model training. The CTC losses of the teacher and student models are also used. }

Specifically, denoting $p_{fp}$, $p_{mp}$ and $p_{i}$ as the output distribution of the full-precision system, mixed-precision quantized system and $i$-bit ($i\in\{2,4,8\}$) uniform-precision quantized system, respectively, the KL regularization term is given by:
\begin{equation}
    \Omega_{i} = D_{KL}(SG(p_{fp})||p_{i});\ 
    \Omega_{mp} = D_{KL}(SG(p_{fp})||p_{mp}),
\end{equation}
where $SG(\cdot)$ denotes the stop-gradient operation preventing the gradient from flowing back to the full-precision systems. 

Let $\mathcal{L}_{fp}$ and $\mathcal{L}_{i}$ denote the CTC losses of the full-precision system and the $i$-bit uniform-precision quantized system ($i \in \{2, 4, 8\}$), respectively. The criterion with KL regularization is given by:
\vspace{-0.1cm}
\begin{equation}
\begin{aligned}
    \mathcal{L}_{kl} = &\mathcal{L}_{fp} + \mathcal{L}_{mp} + \sum_{i\in{\{2,4,8\}}}(\lambda_i\mathcal{L}_{i})\\
    &+ \beta_{mp}\Omega_{mp}  + \beta_{kl}\sum_{i\in{\{2,4,8\}}}(\beta_i\Omega_{i}),
\end{aligned}
\label{eq5}
\end{equation}
where $\lambda_i$, $\beta_{mp}$, $\beta_{kl}$ and $\beta_i$ are constant coefficients.

Note that in Eq.~\ref{eq5}, all the systems share model weights, that is, all the student models are directly quantized from the full-precision teacher. Therefore, combining this with the KL regularization, Eq.~(\ref{Pass 1 eq}) can be modified to $\mathcal{L} = \mathcal{L}{kl} + \eta\cdot \mathcal{C}{size}$.

\section{post-quantization fine-tuning}

So far, we can obtain a well-trained mixed-precision SSL ASR system directly with the proposed Pass 1, all within just one training stage. We further explore the potential for additional performance improvement throughout the second fine-tuning process, which is referred to as \textbf{post-quantization fine-tuning (Pass 2)}.
An overall framework of the proposed method is illustrated in Fig.~\ref{fig1}. For the Pass 2, we study two different initialization approaches: \textbf{1)} freeze the mixed-precision bit-widths searched in Pass 1 and use the same starting point as in Pass 1 for fine-tuning; \textbf{2)} freeze the mixed-precision bit-widths searched in Pass 1 while also utilizing the parameters learned during Pass 1 for initialization. Moreover, the distillation techniques can also be applied in Pass 2.

\section{Experiments}

\vspace{-0.2cm} 
\subsection{Experimental setup}

\noindent\textbf{Baseline systems and data.}
For wav2vec2.0-\textit{base} models, the wav2vec2-base-100h is downloaded from Huggingface\footnote{\href{https://huggingface.co/facebook/wav2vec2-base-100h}{Huggingface: facebook/wav2vec2-base-100h}\label{hf}} as our starting point. We fine-tuned wav2vec2-base-100h for 5 epochs as our baselines. For HuBERT-\textit{large} models, we fine-tuned the pre-trained HuBERT-large-ll60k\footnote{\href{https://huggingface.co/facebook/HuBERT-large-ll60k}{Huggingface: facebook/HuBERT-large-ll60k}} for 20 epochs with a learning rate of 3e-5 as our starting point, and we fine-tuned it for 3 more epochs as our baselines. All systems are trained on LibriSpeech's~\cite{panayotov2015librispeech} 100-hour clean subset. \hnxu{All systems in Pass 1 and Pass 2 were fine-tuned the same number of epochs as their baselines from their starting points.}

\noindent\textbf{Training.}
\textbf{1)} In Pass 1, we utilize the AdamW optimizer with a learning rate of 3e-5 for wav2vec2.0 systems and 1e-5 for HuBERT systems, employing a batch size of around 101 seconds of audio. A linear warmup is implemented for the first 10\% of the training steps, followed by a linear decay to zero. $T$ in the Gumbel-Softmax distribution is annealed from 1 to 0.03. \textbf{2)} In Pass 2, all setups remain the same as in Pass 1, except that the mixed-precision bit-widths is fixed to it searched in Pass 1 and the initialization of model parameters may also be inherited from the learned ones during Pass 1\footnote{For two-stage mixed-precision baselines, Pass 1 refers to the process of training multiple uniform-precision quantized models.}. All experiments are conducted on a single NVIDIA A40 (48 GB).

\noindent\textbf{Quantization.} We apply symmetric quantization on model parameters using quantization-aware training (QAT) with a straight-through estimator and a learnable scale factor~\cite{nagel2021white}. Quantization is primarily applied to Transformer encoder as its massive parameter count; however, to explore the upper limit of the compression ratio, we also quantize the CNNs to 8-bit in a few experiments as mentioned in Sec.~\ref{structure}. 
Therefore, bit-width in tables~\ref{tab1} and~\ref{tab2} refers to the average bit-width of the Transformer encoder only, excluding the other parts.

\vspace{-0.2cm} 
\subsection{Comparison with uniform-precision quantized systems}
We began by comparing mixed-precision quantized systems with uniform-precision quantized systems under the same model-size constraints. For KL regularization in HuBERT systems, when training, we only forward through 3 sub-networks (i.e., the full-precision model and the mixed-precision quantized model, and
another one sampled from the 2-bit, 4-bit and 8-bit models) in one updating iteration instead of all of them to speed up the training process. To facilitate a clearer comparison, we reported
the WER over the entire test set including test clean and test other, as presented in the last column of the tables~\ref{tab1} and~\ref{tab2}, from which we draw the following observations: \textbf{1)} After incorporating KL regularization\footnote{For wav2vec2.0 systems,  \hnxu{$\eta$ and $\beta_{kl}$ are set to $2\times10^{-5}$ and $0.05$ based on WERs of systems O1 and O2 in Tab.~\ref{tab1}, respectively}; For HuBERT systems, \hnxu{$\eta$ and $\beta_{kl}$ are set to $4\times10^{-7}$ and $0.06$ based on WERs of systems O2 and O3 in Tab.~\ref{tab2}, respectively;} Based on
WER of system O2 in Tab.~\ref{tab1}, $\beta_{mp},\beta_2,\beta_4,\lambda_4$ and $\lambda_8$ are set to 1, $\beta_8$ and $\lambda_2$ are set to 0 for all systems.}, the WERs of our one-pass (i.e., performing only Pass 1) mixed-precision quantized systems consistently outperform those of uniform-precision quantized systems. This has been validated for both wav2vec2.0 and HuBERT systems (i.e., systems~O4 v.s. U5 in Tab.~\ref{tab1}; systems~O3 v.s. U5 in Tab.~\ref{tab2}). \textbf{2)} All of our two-pass (i.e., performing Pass 1 and Pass 2) mixed-precision quantized systems demonstrate lower WERs compared to uniform-precision quantized systems of equal or even higher bit-width. Remarkably, the best 4-bit mixed-precision quantized wav2vec2.0 system and 3.8-bit mixed-precision quantized HuBERT system achieve absolute WER reductions of 0.46\% and 0.57\%, respectively, compared to the corresponding 4-bit uniform-precision quantized wav2vec2.0 and HuBERT systems (i.e., systems~F6 v.s. U5 in Tab.~\ref{tab1}; systems~F4 v.s. U5 in Tab.~\ref{tab2}).

\begin{table}[t]
    \centering
     \vspace{-0.3cm}
    \caption{WER($\downarrow$) of wav2vec2.0-base systems with Pass 1 and Pass 2 against 32-bit (bule), uniform-precision quantized (orange) and two-stage mixed-precision quantized (yellow) systems. \hnxu{Prec.: Precision;} Train \hnxu{Uniform Prec.}: Whether to require training uniform-precision quantized models; \hnxu{KL reg.: KL regularization; Weight init.: Where to initialize the weights from; Comp. Ratio: Compression Ratio.} 
     Training time: the total duration of both Pass 1 and Pass 2 (if Pass 2 is included). $\whitediamond$ means the time encompasses the total duration needed to train multiple uniform-precision (i.e., 2-, 4-, 8-bits) quantized models. * means it is significantly better than the full-precision baseline system (U1).  $\dag$ means it has no statistically significant (MAPSSWE~\cite{gillick1989some}, $\alpha$=0.05) WER increase with the full-precision baseline system (U1).}
    \vspace{-0.1cm}
    \setlength\tabcolsep{1pt}
    \resizebox{\linewidth}{!}
    {
    \begin{tabular}{c|c|c|c|c|c|c|c|c|c|cccc|c}
         \hline
          \multirow{4}{*}{ID} & \multirow{4}{*}{\shortstack{Bit-\\width\\(w/o\\CNNs)}} & \multirow{4}{*}{\shortstack{CNNs\\Bit-\\width}} &  \multirow{4}{*}{\shortstack{Mixed\\Prec.}} & \multicolumn{2}{c|}{Pass 1} & \multicolumn{2}{c|}{Pass 2} & \multirow{4}{*}{\shortstack{Comp.\\Ratio}} & \multirow{4}{*}{\shortstack{Training\\Time\\(hrs)}} & \multicolumn{2}{c}{Dev} & \multicolumn{2}{c|}{Test} & \multirow{4}{*}{\shortstack{Test\\all}}\\
         \cline{5-8} \cline{11-14}
         & & & &\multirow{3}{*}{\shortstack{Train\\Uniform\\prec.}} & \multirow{3}{*}{\shortstack{KL\\reg.}} &\multirow{3}{*}{\shortstack{Weight\\init.}}&\multirow{3}{*}{\shortstack{KL\\reg.}} & & & \multirow{3}{*}{clean} & \multirow{3}{*}{other} & \multirow{3}{*}{clean} & \multirow{3}{*}{other}\\
         & & & & & & & & & & & & &\\
         & & & & & & & & & & & & &\\
         \hline

          0&\multicolumn{7}{c|}{$float32$ Starting point}&1.0x & &6.10 	&13.79 &	6.06 &	13.52 &10.00\\
        \hline
         {\cellcolor{cyan!15}U1}&\multicolumn{2}{c|}{32}  &\multirow{7}{*}{\xmark} &\multicolumn{2}{c|}{\multirow{7}{*}{-}} &\multicolumn{2}{c|}{\multirow{7}{*}{-}} &1.0x & 3.8 &5.78	&13.65 &5.89 &	13.32 &9.82 \\
        \cline{1-3} \cline{9-15}
         {\cellcolor{orange!20}U2} &8 & \multirow{5}{*}{32}& &\multicolumn{2}{c|}{} &\multicolumn{2}{c|}{} &3.1x & 4.0 &5.80$^{\dag}$	&13.63$^{\dag}$	&5.83$^{\dag}$ &	13.26$^{\dag}$ &9.76$^{\dag}$\\
         {\cellcolor{orange!20}U3}&6& &&\multicolumn{2}{c|}{}&\multicolumn{2}{c|}{} & \textbf{3.7x} &4.1 &5.86$^{\dag}$ &	13.85$^{\dag}$ &	5.85$^{\dag}$ 	&13.30$^{\dag}$ 	&\textbf{9.79$^{\dag}$} \\
           
          {\cellcolor{orange!20}U4}&5&& &\multicolumn{2}{c|}{}&\multicolumn{2}{c|}{} &4.2x &4.1 & 5.93	&14.08	&5.94$^{\dag}$	&13.73 & 10.06  \\
          {\cellcolor{orange!20}U5}&4&& &\multicolumn{2}{c|}{}&\multicolumn{2}{c|}{} &4.7x & 4.0 &6.12	&14.76	&6.01	&14.31 &10.40\\
          {\cellcolor{orange!20}U6}&2&& &\multicolumn{2}{c|}{}&\multicolumn{2}{c|}{} &6.4x & 4.2 &47.83 &	65.44 &	48.27 &	66.21 &57.75  \\

          \hline
          \hline
          {\cellcolor{yellow!25}M1}&\multirow{2}{*}{4.7}& \multirow{4}{*}{32}&\multirow{4}{*}{\cmark}&\multirow{4}{*}{\cmark}&\multirow{4}{*}{-}&\multirow{4}{*}{ID 0}&\xmark &\multirow{2}{*}{4.3x} &\textbf{17.6$\whitediamond$} &6.32 &	15.27 &	6.36 &	14.54 &	10.68   \\
          {\cellcolor{yellow!25}M2}&&&&&& &\cmark & &28.7$\whitediamond$  &6.14 	&15.07 &	6.08 &	14.45 	&10.50 \\
          \cline{1-2} \cline{8-15}
          {\cellcolor{yellow!25}M3}&\multirow{2}{*}{4.0}&&&&&&\xmark &\multirow{2}{*}{4.7x} &\textbf{17.4$\whitediamond$} &6.41&	15.65& 	6.39 	&14.86 &10.87\\
          {\cellcolor{yellow!25}M4}&&&&&& &\cmark & &28.9$\whitediamond$ & 6.15 &	15.40 &	6.30 	&14.63 &10.70\\
          \hline
          \hline
          O1&\multirow{2}{*}{4.6}& \multirow{4}{*}{32}&\multirow{4}{*}{\cmark}&\multirow{4}{*}{\xmark}&\multirow{1}{*}{\xmark}&\multicolumn{2}{c|}{\multirow{4}{*}{\xmark}} &\multirow{2}{*}{4.4x} &5.2 &6.60	&15.88	&6.67	&15.41 &11.29 \\
           O2&&&&&\multirow{1}{*}{\cmark} &\multicolumn{2}{c|}{} & &21.5 &5.80$^{\dag}$ 	&14.45 	&5.97$^{\dag}$ 	&13.88 &10.15 \\
           \cline{1-2} \cline{6-6} \cline{9-15}
           O3&\multirow{2}{*}{4.0}&&&&\multirow{1}{*}{\xmark}&\multicolumn{2}{c|}{} &\multirow{2}{*}{4.7x} &5.3 &7.08 	&16.04 	&7.07 	&15.61 &11.59  \\
           O4&&&&&\multirow{1}{*}{\cmark} &\multicolumn{2}{c|}{} &&21.2 &5.94 	&14.65 	&5.95$^{\dag}$ 	&14.19 & 10.31 \\
          \hline
          \hline
          N1&\multirow{4}{*}{4.6}& \multirow{8}{*}{32}&\multirow{8}{*}{\cmark}&\multirow{8}{*}{\xmark}&\multirow{2}{*}{\xmark}&\multirow{8}{*}{ID 0}&\xmark &\multirow{4}{*}{4.4x} &\textbf{9.2} &5.94	&14.11	&5.97$^{\dag}$	&13.69 &10.05\\
          N2&&&&&&&\cmark & &21.3 &5.69$^{\dag}$& 	14.39 	&5.75* &	13.59 &9.89$^{\dag}$ \\
          \cline{6-6} \cline{8-8}
          N3&&&&&\multirow{2}{*}{\cmark}&&\xmark & &25.8 &5.96 	&14.53 &	6.05 &	13.88  &10.19\\
          N4&&&&&& &\cmark & &37.6 &5.71$^{\dag}$ 	&14.24 	&5.79$^{\dag}$ 	&13.60 &9.92$^{\dag}$\\ 
          
          \cline{1-2} \cline{6-6} \cline{8-15}
          N5&\multirow{4}{*}{4.0}&&&&\multirow{2}{*}{\xmark}&&\xmark & \multirow{4}{*}{4.7x}&\textbf{9.4} &6.05 	&14.70 	&6.11 	&14.07 &10.32\\
          N6&&&&&&&\cmark & &21.3 &5.77$^{\dag}$ 	&14.40 &	5.86$^{\dag}$ &	13.79 &10.05\\
          \cline{6-6} \cline{8-8}
          N7&&&&&\multirow{2}{*}{\cmark}&&\xmark & &25.3 &5.98 	&14.73 	&6.12 	&13.93 &10.25 \\
          N8&&&&&& &\cmark & &37.5 &5.75$^{\dag}$ 	&14.45 	&5.81$^{\dag}$ 	&13.92  &10.10\\ 
          \hline
          \hline
          F1&\multirow{4}{*}{4.6}& \multirow{8}{*}{32}&\multirow{8}{*}{\cmark}&\multirow{8}{*}{\xmark}&\multirow{2}{*}{\xmark}&\multirow{8}{*}{Pass 1}&\xmark &\multirow{4}{*}{4.4x} &9.4 &5.87$^{\dag}$ 	&14.84 	&6.07 	&14.41 &10.48\\
          F2&&&&&&&\cmark & &21.4 &5.74$^{\dag}$ &	14.44 	&5.64* 	&13.59 &\textbf{9.84$^{\dag}$}\\
          \cline{6-6} \cline{8-8}
          F3&&&&&\multirow{2}{*}{\cmark}&&\xmark & &25.8 &5.75$^{\dag}$ 	&14.53 	&5.88$^{\dag}$ 	&14.06 &10.20\\
          F4&&&&&& &\cmark & &37.7 &5.53* 	&14.26 &	5.62* 	&13.66 &9.87$^{\dag}$\\
          \cline{1-2} \cline{6-6} \cline{8-8} \cline{9-15}
          F5&\multirow{4}{*}{4.0}&&&&\multirow{2}{*}{\xmark}&&\xmark &\multirow{4}{*}{\textbf{4.7x}} &9.3  & 6.07 	& 15.00 & 	6.09 & 	14.14 &10.35\\
          F6&&&&&&&\cmark & &21.4 &5.77$^{\dag}$ &	14.39 	&5.72*	&13.71 &\textbf{9.94$^{\dag}$} \\
          \cline{6-6} \cline{8-8}
          F7&&&&&\multirow{2}{*}{\cmark}&&\xmark & &25.3 &5.80$^{\dag}$ 	&14.59 	&5.91$^{\dag}$ 	&14.11 &10.25 \\
          F8&&&&&& &\cmark & &37.7 &5.63* 	&14.28 &	5.77$^{\dag}$ 	&13.85 &10.04\\
          \hline
          F9&4.6&\multirow{1}{*}{8}&\multirow{1}{*}{\cmark}&\multirow{1}{*}{\xmark}&\multirow{1}{*}{\xmark}&\multirow{1}{*}{Pass 1} &\cmark &\textbf{6.4x} &22.3 &5.67$^{\dag}$ &	14.15 &	5.81$^{\dag}$ &	13.57 &	\textbf{9.91$^{\dag}$} \\
          \hline

    \end{tabular}
    }
    \label{tab1}
    \vspace{-0.5cm}
\end{table}

\begin{table}[t]
    \centering
    \vspace{-0.3cm}
    \caption{WER($\downarrow$) of HuBERT-large systems with Pass 1 and Pass 2 against 32-bit, uniform-precision quantized and two-stage mixed-precision quantized systems. The headings and marks have the same meanings as those in Table~\ref{tab1}.}
    \vspace{-0.1cm}
    \setlength\tabcolsep{1pt}
    \resizebox{\linewidth}{!}
    {
    \begin{tabular}{c|c|c|c|c|c|c|c|c|c|cccc|c}
         \hline
          \multirow{4}{*}{ID} & \multirow{4}{*}{\shortstack{Bit-\\width\\(w/o\\CNNs)}} & \multirow{4}{*}{\shortstack{CNNs\\Bit-\\width}} &\multirow{4}{*}{\shortstack{Mixed\\Prec.}} & \multicolumn{2}{c|}{Pass 1} & \multicolumn{2}{c|}{Pass 2} & \multirow{4}{*}{\shortstack{Comp.\\Ratio}} & \multirow{4}{*}{\shortstack{Training\\Time\\(hrs)}} & \multicolumn{2}{c}{Dev} & \multicolumn{2}{c|}{Test} & \multirow{4}{*}{\shortstack{Test\\all}}\\
         \cline{5-8} \cline{11-14}
         & && &\multirow{3}{*}{\shortstack{Train\\Uniform\\Prec.}} & \multirow{3}{*}{\shortstack{KL\\reg.}} &\multirow{3}{*}{\shortstack{Weight\\init.}}&\multirow{3}{*}{\shortstack{KL\\reg.}} & & & \multirow{3}{*}{clean} & \multirow{3}{*}{other} & \multirow{3}{*}{clean} & \multirow{3}{*}{other}\\
         & & & & & & & & & & & & &\\
         & & & & & & & & & & & & &\\
         \hline

          0&\multicolumn{7}{c|}{$float32$ Starting point}&1.0x & &3.98 &	8.48 &	4.09 	&8.45 &	6.40 \\
        \hline
         {\cellcolor{cyan!15}U1}&\multicolumn{2}{c|}{32}&\multirow{6}{*}{\xmark} &\multicolumn{2}{c|}{\multirow{6}{*}{-}} &\multicolumn{2}{c|}{\multirow{6}{*}{-}} &1.0x &4.1 &3.93 &	8.33 &	4.00& 	8.40 &	6.33 \\
         \cline{1-3} \cline{9-15}
         {\cellcolor{orange!20}U2} &8&\multirow{5}{*}{32}& &\multicolumn{2}{c|}{} &\multicolumn{2}{c|}{} &3.5x &6.4 
           &3.91$^{\dag}$ &	8.38$^{\dag}$ &	4.02$^{\dag}$ 	&8.39$^{\dag}$  	&6.33$^{\dag}$ \\

        {\cellcolor{orange!20}U3} &6&& &\multicolumn{2}{c|}{} &\multicolumn{2}{c|}{} & \textbf{4.5x} &6.7 &3.93$^{\dag}$ &	8.44 &	4.01$^{\dag}$ 	&8.47$^{\dag}$ 	&\textbf{6.36$^{\dag}$} \\
          {\cellcolor{orange!20}U4}&5&& &\multicolumn{2}{c|}{}&\multicolumn{2}{c|}{} &5.2x &6.6 &4.01 	&8.64 	&4.06$^{\dag}$ &	8.69	&6.51 \\  
          
          {\cellcolor{orange!20}U5}&4&& &\multicolumn{2}{c|}{}&\multicolumn{2}{c|}{} &6.2x &6.7  &4.17 	&9.26 &	4.14 &	8.99 &	6.70\\
          {\cellcolor{orange!20}U6}&2&& &\multicolumn{2}{c|}{}&\multicolumn{2}{c|}{} &9.8x &6.4 &6.86 &	16.67 &	7.04 &	16.62&	12.10 \\

            
          \hline
          \hline
          {\cellcolor{yellow!25}M1}&\multirow{2}{*}{3.8}&\multirow{2}{*}{32}&\multirow{2}{*}{\cmark}&\multirow{2}{*}{\cmark}&\multirow{2}{*}{-}&\multirow{2}{*}{ID 0}&\xmark &\multirow{2}{*}{6.4x} &\textbf{24.8$\whitediamond$} &4.11 &	9.20 &	4.16 	&8.92 &	6.68 \\
          {\cellcolor{yellow!25}M2}&&&&&& &\cmark & &31.2$\whitediamond$ &3.96$^{\dag}$ &	9.21 &	4.01$^{\dag}$ 	&8.96 	&6.63 \\
          \hline
          \hline
          O1&4.0&\multirow{4}{*}{32}&\multirow{4}{*}{\cmark}&\multirow{4}{*}{\xmark}&\cmark&\multicolumn{2}{c|}{\multirow{4}{*}{\xmark}} &\textbf{6.2x} &32.9 &3.84$^{\dag}$ 	&8.58 	&3.91$^{\dag}$ &	8.41$^{\dag}$  &\textbf{6.29$^{\dag}$}\\
          \cline{1-2} \cline{6-6} \cline{9-15}
          O2&\multirow{2}{*}{3.8}&&&&\multirow{1}{*}{\xmark} &\multicolumn{2}{c|}{} &\multirow{2}{*}{6.4x} &11.3 &4.34 	&10.12 &	4.42 &	9.80 &	7.26 \\

           O3&&&&&\multirow{1}{*}{\cmark} &\multicolumn{2}{c|}{} & &31.1  &	3.91$^{\dag}$ &	9.16 &	3.97$^{\dag}$ &	9.02 &	6.64 \\
           \cline{1-2} \cline{6-6} \cline{9-15}
           O4&3.5&&&&\multirow{1}{*}{\cmark} &\multicolumn{2}{c|}{} &6.8x &31.5 &4.01 &	9.69 &	4.04 &	9.34 &6.84 \\

          \hline
          \hline

          N1&\multirow{4}{*}{3.8}&\multirow{4}{*}{32}&\multirow{4}{*}{\cmark}&\multirow{4}{*}{\xmark}&\multirow{2}{*}{\xmark}&\multirow{4}{*}{ID 0}&\xmark &\multirow{4}{*}{6.4x} &\textbf{16.6}  & 4.00$^{\dag}$ &	9.22 &	4.03$^{\dag}$ &	8.88 &	6.59 \\

          N2&&&&&&&\cmark & &23.6 &3.91$^{\dag}$ &	9.14 &	3.91$^{\dag}$ 	&8.80 &	6.50\\
          \cline{6-6} \cline{8-8}

          N3&&&&&\multirow{2}{*}{\cmark}&&\xmark & &36.5 &3.98$^{\dag}$ &	8.95 &	4.03$^{\dag}$ 	&8.91 &	6.61 \\

          N4&&&&&& &\cmark & &44.7 &3.89$^{\dag}$ 	&8.99 &	3.90$^{\dag}$ 	&8.74 	&6.46 \\ 
          \hline
          \hline
          F1&\multirow{4}{*}{3.8}&\multirow{6}{*}{32}&\multirow{6}{*}{\cmark}&\multirow{6}{*}{\xmark}&\multirow{2}{*}{\xmark}&\multirow{6}{*}{Pass 1}&\xmark &\multirow{4}{*}{6.4x} &16.7 &3.92$^{\dag}$ &	9.10 	&3.97$^{\dag}$ &	8.77 &6.51  \\
          F2&&&&&&&\cmark & &23.7 & 3.84$^{\dag}$ 	&8.94 	&3.85* 	&8.60	&6.36$^{\dag}$\\
          \cline{6-6} \cline{8-8}
          F3&&&&&\multirow{2}{*}{\cmark}&&\xmark & &36.6 &3.77* 	&8.50 	&3.83* 	&8.42$^{\dag}$ 	&6.26$^{\dag}$ \\
          F4&&&&&& &\cmark & &43.0
          &3.74* &	8.41$^{\dag}$ 	&3.70* &	8.29$^{\dag}$  &	\textbf{6.13*} \\
          \cline{1-2} \cline{6-6} \cline{8-15}
          F5&\multirow{2}{*}{3.5}&&&&\multirow{2}{*}{\cmark}&&\xmark &\multirow{2}{*}{\textbf{6.8x}} &36.8 &3.88$^{\dag}$ &	8.89 &	3.90$^{\dag}$ 	&8.77 &6.47\\
          F6&&&&&& &\cmark & &43.8 &3.83$^{\dag}$ &	8.91 &	3.77* 	&8.65 &\textbf{6.35$^{\dag}$}\\
          \hline
          F7&3.5&8&\multirow{1}{*}{\cmark}&\multirow{1}{*}{\xmark}&\multirow{1}{*}{\cmark}&\multirow{1}{*}{Pass 1} &\cmark &\textbf{8.6x} &43.9 &3.84$^{\dag}$ &	8.97 &	3.78* &	8.70 &	\textbf{6.38$^{\dag}$} \\
          
          \hline

    \end{tabular}
    }
    \label{tab2}
    \vspace{-0.5cm}
\end{table}

\vspace{-0.2cm} 
\subsection{Comparison with two-stage mixed-precision quantized systems}

Under the same initialization and training settings, the WERs of our two-pass mixed-precision quantized systems are consistently lower than those of traditional two-stage mixed-precision quantized systems~\cite{mq-jh} (e.g., systems~N1, N3 v.s. system~M1; systems~N2, N4 v.s. system~M2; systems~N5, N7 v.s. system~M3 and systems~N6, N8 v.s. system~M4 in Tab.~\ref{tab1}, respectively; systems~N1, N3 v.s. system~M1 and systems~N2, N4 v.s. system~M2 in Tab.~\ref{tab2}, respectively). This indicates that the proposed joint mixed-precision learning and quantized model training stage can effectively learn a better mixed-precision bit-widths compared to the inconsistent two-stage mixed-precision method. For example, the wav2vec2.0 system of N1 exhibits an absolute WER reduction of 0.63\% against system~M1, accompanied by a training time speed-up ratio of 1.9x in Tab.~\ref{tab1}. Likewise, a training time speed-up ratio of 1.5x for the HuBERT system of system~N1 is observed compared to system~M1 in Tab.~\ref{tab2}. 
Notably, with KL regularization, the proposed Pass 1 alone can produce mixed-precision quantized wav2vec2.0 systems with WERs even lower than those of two-stage mixed-precision systems (i.e., systems~O2, O4 v.s. systems~M2, M4 in Tab.~\ref{tab1}, respectively) as well as a 4-bit lossless mixed-precision quantized HuBERT system.

\vspace{-0.2cm} 
\subsection{Comparison with 32-bit full-precision systems}

When performing uniform-precision quantization, experiments indicate that lossless compression can only be achieved with quantization of no less than 6-bit, leading to a maximum compression ratio of 3.7x for the wav2vec2.0 system and 4.5x for the HuBERT systems, respectively (i.e., system~U3 in Tab.~\ref{tab1} and system~U3 in ~\ref{tab2}). In contrast, we finally obtain a 4-bit mixed-precision quantized wav2vec2.0 system and a 3.5-bit mixed-precision quantized HuBERT system with lossless compression ratios of 4.7x and 6.8x (i.e., system~F6 in Tab.~\ref{tab1} and system~F6 in Tab.~\ref{tab2}), respectively.

To explore the upper limit of the compression ratio, we further quantized the CNNs with 8 bits, resulting in a 4.6-bit mixed-precision quantized wav2vec2.0 system and a 3.5-bit mixed-precision quantized HuBERT system that achieved a compression ratio of \textbf{6.4x} and \textbf{8.6x}, respectively, also without statistically significant WER increase. A comparison of WERs on LibriSpeech dev clean\footnote{Only the WER on dev clean is presented as the WERs on other subsets are not reported in ~\cite{peng2021shrinking} and ~\cite{mp-w2v}.} between previously published SSL ASR systems and our system is shown in Tab.~\ref{tab3}.

\begin{table}[h]
    \centering
    \vspace{-0.3cm}
    \caption{Absolute (abs.) and relative (rel.)WER($\downarrow$) reduction on Librispeech dev clean compared to their respective baselines between published systems and ours. * means it is not provided directly and is calculated by us. }
    \vspace{-0.1cm}
    \setlength\tabcolsep{1pt}
    \resizebox{\linewidth}{!}
    {
    \begin{tabular}{c|c|c|c}
         \hline
           \multirow{2}{*}{System} & \multirow{2}{*}{{\shortstack{Model\\Size}}} & \multirow{2}{*}{{\shortstack{Comp.\\Ratio}}} & Dev clean\\
         & & &(abs. $\downarrow$ / rel. $\downarrow$)\\
         \hline
         8-bit uniform-precision wav2vec2.0-large~\cite{peng2021shrinking} &\multirow{4}{*}{1262MB} &3.6x &2.75(+0.12/4.6\%) \\
         4-bit uniform-precision + 50\% Sparse wav2vec2.0-large~\cite{mp-w2v} & &9.8x*  &4.65(+0.45/10.7\%)  \\
         4-bit uniform-precision wav2vec2.0-large~\cite{mp-w2v} & &6.2x*  &4.53(+0.33/7.9\%) \\
         \textbf{3.5-bit mixed-precision HuBERT-large (system F7 in Tab.~\ref{tab2}, ours)} & & 8.6x  &3.84\textbf{(-0.09/-2.3\%)} \\
         \hline
         6-bit UP + 1 encoder layer dropped wav2vec2.0-base~\cite{onepass-zq} &\multirow{2}{*}{378MB} & 3.9x  &5.77(-0.01/0\%)\\
         \textbf{4.6-bit MP wav2vec2.0-base (system F9 in Tab.~\ref{tab1}, ours)} & & \textbf{6.4x} &\textbf{5.67(-0.11/-2\%)} \\
         \hline

    \end{tabular}
    }
    \label{tab3}
    \vspace{-0.3cm}
\end{table}

\section{Conclusion}
We propose a novel joint mixed-precision learning and quantized model training method with an optional post-quantization fine-tuning process for SSL speech foundation models. Our system enables maintaining the consistency between mixed-precision learning and quantized model training, as well as recovering performance with just a few steps of fine-tuning. Extensive results demonstrate that under the same model-size (bit-width) constraints, our systems outperform the uniform-precision quantized systems and the two-stage mixed-precision quantized systems. 

\vspace{-0.15cm}
\section*{Acknowledgment}
This research is supported by Hong Kong RGC GRF grant No. 14200220, 14200021, 14200324 and Innovation Technology Fund grant No. ITS/218/21.

\label{sec:refs}


\bibliographystyle{IEEEtran}
\bibliography{reference}


\end{document}